\documentclass[prl,twocolumn,showpacs,floatfix,superscriptaddress,amsmath,amsfonts,amssymb,preprintnumbers,citeautoscript]{revtex4}
\usepackage[T1]{fontenc}\usepackage[latin1]{inputenc}
\usepackage{dcolumn,graphicx,color,booktabs,microtype,afterpage}
\usepackage[charter]{mathdesign}
\usepackage{sidecap}
\usepackage{ulem}
\renewcommand{\figurename}{Fig.}

\makeatletter\renewcommand{\fnum@figure}[1]{\figurename~\thefigure~(color online).}\makeatother

\usepackage[colorlinks,plainpages=false,linkcolor=blue,urlcolor=blue,citecolor=black,pdfpagemode=UseNone,pdfstartview=FitBH]{hyperref}

\begin{document}

\title{Orbital control of noncollinear magnetic order in nickelate heterostructures}

\author{A.~Frano}
\affiliation{Max-Planck-Institut~f\"{u}r~Festk\"{o}rperforschung, Heisenbergstr.~1, D-70569 Stuttgart, Germany}
\affiliation{Helmholtz-Zentrum Berlin f\"{u}r Materialien und Energie, Wilhelm-Conrad-R\"{o}ntgen-Campus BESSY II, Albert-Einstein-Str. 15, D-12489 Berlin, Germany}

\author{E.~Schierle}
\affiliation{Helmholtz-Zentrum Berlin f\"{u}r Materialien und Energie, Wilhelm-Conrad-R\"{o}ntgen-Campus BESSY II, Albert-Einstein-Str. 15, D-12489 Berlin, Germany}

\author{M.~W.~Haverkort}
\affiliation{Max-Planck-Institut~f\"{u}r~Festk\"{o}rperforschung, Heisenbergstr.~1, D-70569 Stuttgart, Germany}
\affiliation{Quantum Matter Institute, University of British Columbia, Vancouver, B.C. V6T 1Z1, Canada}

\author{Y.~Lu}
\affiliation{Max-Planck-Institut~f\"{u}r~Festk\"{o}rperforschung, Heisenbergstr.~1, D-70569 Stuttgart, Germany}

\author{M.~Wu}
\affiliation{Max-Planck-Institut~f\"{u}r~Festk\"{o}rperforschung, Heisenbergstr.~1, D-70569 Stuttgart, Germany}

\author{S.~Blanco-Canosa}
\affiliation{Max-Planck-Institut~f\"{u}r~Festk\"{o}rperforschung, Heisenbergstr.~1, D-70569 Stuttgart, Germany}

\author{U.~Nwankwo}
\affiliation{Max-Planck-Institut~f\"{u}r~Festk\"{o}rperforschung, Heisenbergstr.~1, D-70569 Stuttgart, Germany}

\author{A.~V.~Boris}
\affiliation{Max-Planck-Institut~f\"{u}r~Festk\"{o}rperforschung, Heisenbergstr.~1, D-70569 Stuttgart, Germany}

\author{P.~Wochner}
\affiliation{Max-Planck-Institut~f\"{u}r~Festk\"{o}rperforschung, Heisenbergstr.~1, D-70569 Stuttgart, Germany}

\author{G.~Cristiani}
\affiliation{Max-Planck-Institut~f\"{u}r~Festk\"{o}rperforschung, Heisenbergstr.~1, D-70569 Stuttgart, Germany}

\author{H.~U.~Habermeier}
\affiliation{Max-Planck-Institut~f\"{u}r~Festk\"{o}rperforschung, Heisenbergstr.~1, D-70569 Stuttgart, Germany}

\author{G.~Logvenov}
\affiliation{Max-Planck-Institut~f\"{u}r~Festk\"{o}rperforschung, Heisenbergstr.~1, D-70569 Stuttgart, Germany}

\author{V.~Hinkov}
\affiliation{Quantum Matter Institute, University of British Columbia, Vancouver, B.C. V6T 1Z1, Canada}

\author{E.~Benckiser}
\affiliation{Max-Planck-Institut~f\"{u}r~Festk\"{o}rperforschung, Heisenbergstr.~1, D-70569 Stuttgart, Germany}

\author{E.~Weschke}
\affiliation{Helmholtz-Zentrum Berlin f\"{u}r Materialien und Energie, Wilhelm-Conrad-R\"{o}ntgen-Campus BESSY II, Albert-Einstein-Str. 15, D-12489 Berlin, Germany}

\author{B.~Keimer}
\email[]{b.keimer@fkf.mpg.de}
\affiliation{Max-Planck-Institut~f\"{u}r~Festk\"{o}rperforschung, Heisenbergstr.~1, D-70569 Stuttgart, Germany}

\begin{abstract}
We have used resonant x-ray diffraction to develop a detailed description of antiferromagnetic ordering in epitaxial superlattices based on two-unit-cell thick layers of the strongly correlated metal LaNiO$_3$. We also report reference experiments on thin films of PrNiO$_3$ and NdNiO$_3$. The resulting data indicate a spiral state whose polarization plane can be controlled by adjusting the Ni $d$-orbital occupation via two independent mechanisms: epitaxial strain and quantum confinement of the valence electrons. The data are discussed in the light of recent theoretical predictions.
\end{abstract}


\maketitle

The theoretical prediction and experimental control of collective electronic ordering phenomena in metals are among the greatest challenges of current solid-state physics. Recent advances in the synthesis of epitaxial metal-oxide heterostructures have provided new opportunities for the exploration and manipulation of the phase behavior of metallic conductors \cite{Mannhart2010,Hwang2012}. However, detecting the subtle spin, charge, and orbitally modulated structures resulting from Fermi surface instabilities in the atomically thin layers of such devices remains challenging. Most experiments to-date have focused on spatially uniform order parameters (such as the uniform magnetization) that modify the macroscopic properties, and on corresponding layer-to-layer variations which can be determined by neutron or x-ray reflectometry. In contrast, little experimental information is currently available about ordering patterns within the heterostructure layers, despite many predictions in this regard.

We have used resonant x-ray diffraction (RXD) to develop a detailed, microscopic description of the magnetic ordering patterns in epitaxial superlattices (SLs) of metallic perovskites LaNiO$_3$ (LNO)and thin films of $R$NiO$_3$ ($R$ = Nd and Pr). Bulk analogs of these materials either remain paramagnetic at all temperatures ($R$ = La), or become Mott-insulating and antiferromagnetic at low temperatures ($R \neq$ La) \cite{Torrance1992,Garcia1992}. Recent experimental \cite{Boris2011,Benckiser2011,Scherwitzl2011,Freeland2011,Chakhalian2011,Gibert2012} and theoretical \cite{Chaloupka2008,Hansmann2009,Hansmann2010,Han2010,Han2011,LeePRL2011,LeePRB2011,Dong2013,Lau2013} work has focused on the phase behavior of $R$NiO$_3$-based thin films and heterostructures. In particular, muon spin rotation ($\mu$SR) experiments yielded evidence of antiferromagnetism in SLs with LNO layers of thickness below three unit cells (u.c.) sandwiched between insulating blocking layers, whereas SLs with thicker LNO layers remain paramagnetic \cite{Boris2011}.  Exchange-bias effects revealed by magnetometric measurements on heterostructures of LNO and ferromagnetic LaMnO$_3$ also indicated antiferromagnetic or spin-glass order in LNO \cite{Gibert2012}.  These results have established the nickelates as a promising platform for phase control of metals in proximity to a Mott transition, and they have stimulated detailed theoretical predictions for the magnetic ordering patterns in nickelate heterostructures and SLs \cite{LeePRL2011,LeePRB2011,Dong2013,Lau2013}.  In analogy to prior work on bulk-like NdNiO$_3$ (NNO) films \cite{Scagnoli2006},  we have used RXD to show that the Ni moments in the SLs form a spiral state. We further demonstrate control of the polarization plane of the spiral by manipulating the Ni $d$-orbital occupation via epitaxial strain and quantum confinement. Our RXD data enable detailed tests of theoretical work on the nickelates, and provide the foundation for the integration of ``orbitally engineered'' nickelates into spintronic devices.

\begin{figure}[b]
\includegraphics[width=0.99\columnwidth]{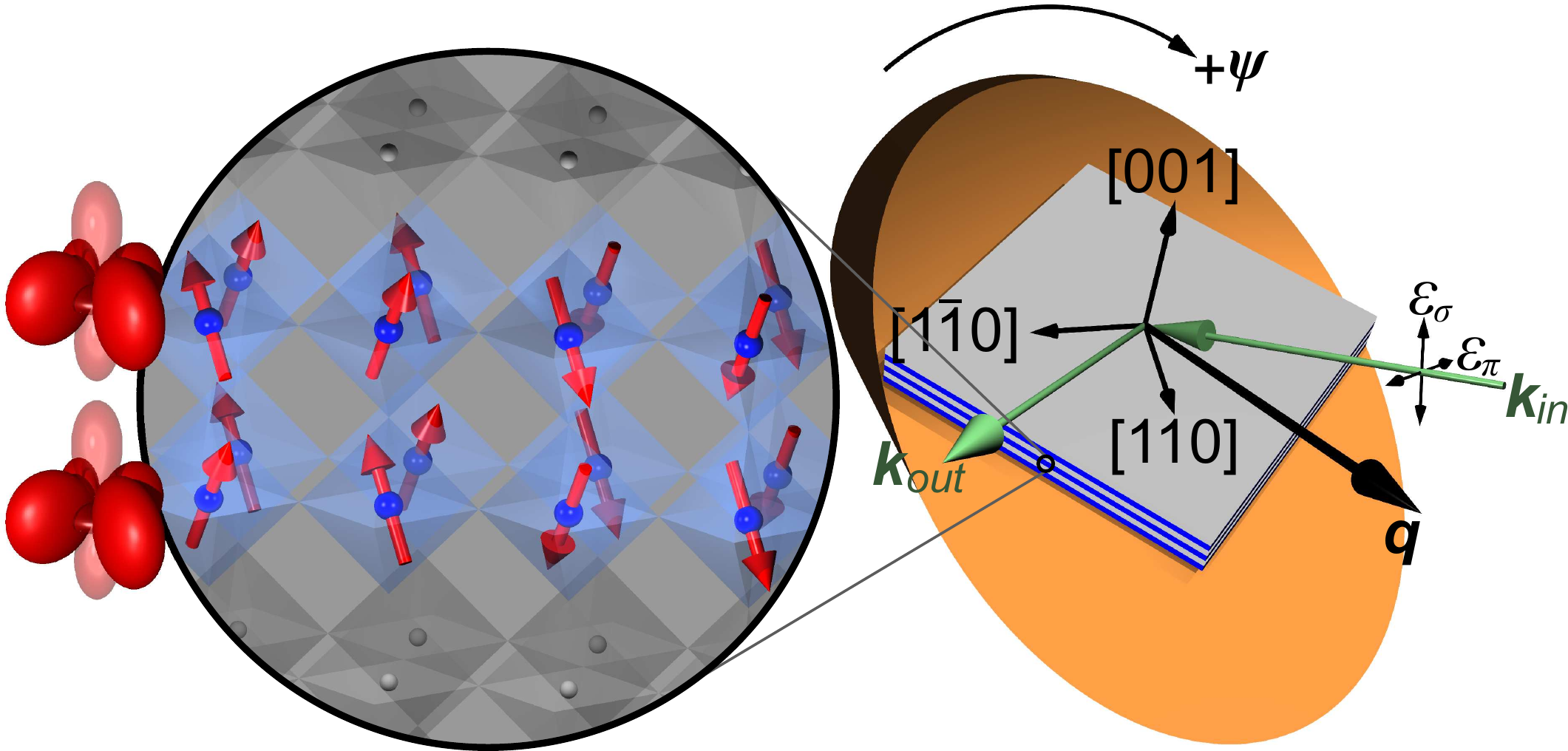}
\caption{Left panel: Schematic diagram of the magnetic structure of a $N=2$ LNO-RXO SL grown on tensile strain. The Ni moments are illustrated by red arrows. Also shown are the electron density distributions in the Ni $d_{x^2-y^2}$ (opaque) and $d_{3z^2-r^2}$ (semitransparent) orbitals. Right panel: samples were mounted on a tilted wedge to access the $[111]$ direction of the pseudocubic perovskite structure in the horizontal scattering geometry. Also shown are the wave vectors of the incoming and outgoing photons (green arrows), the corresponding momentum transfer $\textbf{q}$, the incoming photon polarization $\varepsilon$ (black arrows), and the azimuthal angle $\psi$.}
\label{fig:figure1}
\end{figure}

\begin{figure*}[t]
\includegraphics[width=1.95\columnwidth]{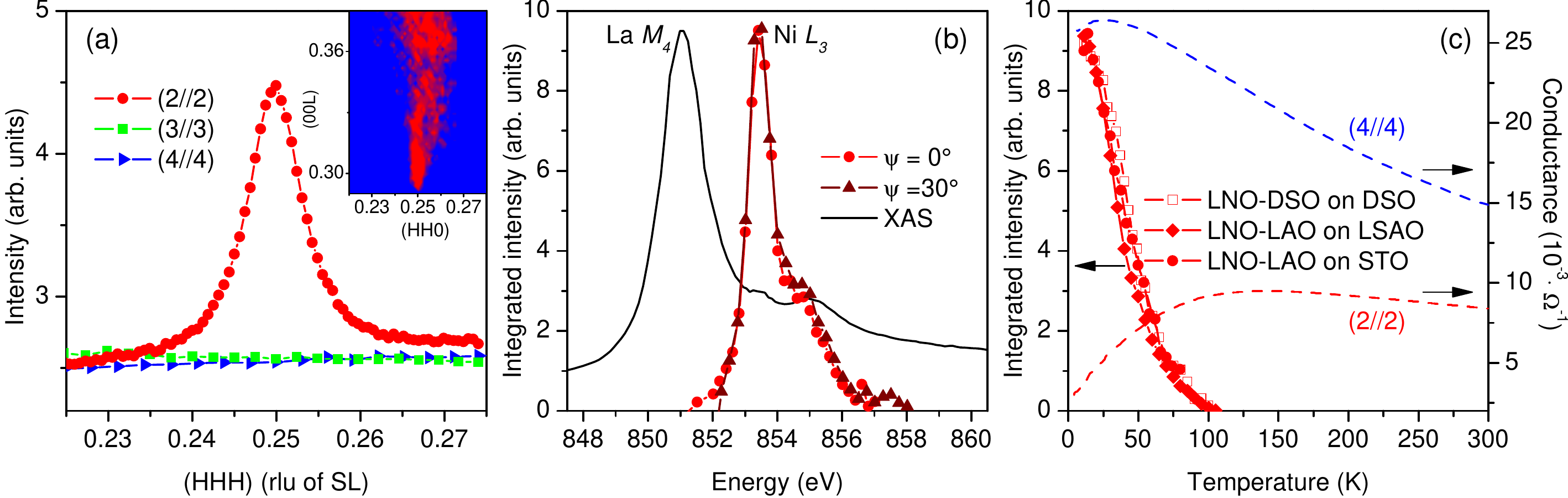}
\caption{(a) Scans around $\textbf{q}_{SDW}=(\frac{1}{4},\frac{1}{4},L)$ for SLs with $N=2$ consecutive LNO unit cells show a magnetic Bragg reflection, while those with $N=3,4$ do not. The inset shows a reciprocal-space map of the scattered intensity from the N=2 SL. (b) Comparison between the x-ray absorption spectrum (XAS) and the photon-energy dependence of the magnetic Bragg intensity at $\textbf{q}_{SDW}$ close to the Ni $L_3$ ($2p\rightarrow 3d$) edge, which shows almost identical line shapes for two azimuthal angles $\psi=0, 30^{\circ}$. (c) Temperature dependence of the magnetic Bragg intensity at $\textbf{q}_{SDW}$ in LNO-based SLs with $N=2$ (red symbols), compared to representative $dc$ electrical conductance measurements (taken by a standard four-probe method) for $N=2$ and $N=4$ SLs.}
\label{fig:figure2}
\end{figure*}

Superlattices of LNO and insulating, nonmagnetic LaAlO$_3$ (LAO) or DyScO$_3$ (DSO) \cite{DyScO3} were grown using pulsed laser deposition on various [001]-oriented substrates: La(Sr)AlO$_4$ (LSAO), [LaAlO$_{3}$]$_{0.3}$[Sr$_{2}$AlTaO$_6$]$_{0.7}$ (LSAT), SrTiO$_3$ (STO), and DSO with in-plane lattice constants 3.756, 3.87, 3.905, and  3.95 {\AA}, respectively. The substrates and blocking layers allowed us to impose variable levels of compressive or tensile strain on LNO, whose bulk lattice constant is $a =3.84$ {\AA}. Data on NNO and PrNiO$_3$ (PNO) films with $a =3.79$ and $3.84$ {\AA}, respectively, were taken for comparison. We discuss structural data in terms of the pseudocubic unit cell of the perovskite lattice. The momentum transfer $\textbf{q} = (H,K,L)$ is also indexed in these units. As demonstrated previously \cite{Boris2011,Benckiser2011}, the layer sequence of our SLs is defined with u.c. precision. We consider SLs with $N$\,u.c. LNO - $N$\,u.c. \textit{R}XO ($N=2,3,4$) structures, labeled $N$//$N$) in the following.

The RXD experiments were performed at the BESSY-II undulator beam line UE46-PGM1 using variable linearly polarized photons in a horizontal scattering geometry (Fig.~\ref{fig:figure1}). A three-circle UHV diffractometer was equipped with a continuous-flow He cryostat. To access the $(\frac{1}{4},\frac{1}{4},\frac{1}{4})$ reflection, the [001]-oriented samples were mounted on copper wedges with a 55$^{\circ}$ tilt. The azimuth value $\psi=0^{\circ}$ is assigned when the crystal vectors $[1,1,1]$ and $[1,\overline{1},0]$ span the scattering plane, and positive rotation around $\textbf{q}$ is left-handed.  This geometry precluded full $\psi$-scans, as there are two angular ranges where the incident or outgoing beams are below the sample horizon. In our definition of $\psi$, these occur around $\psi=270^{\circ}$ and $90^{\circ}$, respectively. To estimate the values of the polarization-dependent scattering cross section at these azimuth positions, we used a non-tilted (asymmetric) scattering geometry to access part of the $(\frac{1}{4},\frac{1}{4},L)$ scattering rod.

Figure~\ref{fig:figure2}(a) shows selected reciprocal-space scans on (2//2), (3//3), and (4//4) SLs taken under resonant conditions, that is, with the photon energy tuned to the Ni $L_3$ edge. (2//2) SLs of {\it all} compositions grown on {\it all} substrates investigated here show resonant Bragg reflections at $\textbf{q} = (\frac{1}{4},\frac{1}{4},L)$, whereas samples with 3 or 4 consecutive u.c. of LNO do not. Comparison of the energy dependence of the diffracted intensity at this $\textbf{q}$ with x-ray absorption spectroscopy (XAS) data (Fig.~\ref{fig:figure2}(b)) confirms the resonance at the Ni $L_3$ edge, and the polarization dependence of the scattering cross section (see below) confirms the magnetic origin of the Bragg reflections. The RXD data thus yield the ordering vector of the antiferromagnetic state identified by $\mu$SR \cite{Boris2011}. The observation of magnetic Bragg reflections implies staggered ordering of the Ni moments in the SL plane analogous to the one characterizing antiferromagnetic order in bulk nickelate perovskites \cite{Garcia1992,Scagnoli2006}
and rules out a spin glass state that has been considered in related SLs \cite{Gibert2012}. The observation of a scattering rod perpendicular to the SL plane (inset in Fig.~\ref{fig:figure2}(a)) is consistent with quasi-two-dimensional order, but the accessible range of the corresponding momentum transfer, $L$, is not sufficient to determine the spin-spin correlation length in this direction.
Figure~\ref{fig:figure2}(c) demonstrates that the onset of the magnetic Bragg intensity as a function of temperature coincides with the maximum observed in $dc$ conductance measurements of $N=2$ SLs. This observation agrees with recent predictions according to which the magnetic order in LNO-based SLs is a consequence of spin density wave (SDW) formation  \cite{LeePRL2011,LeePRB2011}. We thus refer to the magnetic ordering vector as $\textbf{q}_{SDW}$. Further work is required to elucidate whether this transition is accompanied by a modification of the charge density \cite{Scagnoli2005}.

\begin{figure*}[t]
\center\includegraphics[width=0.98\textwidth]{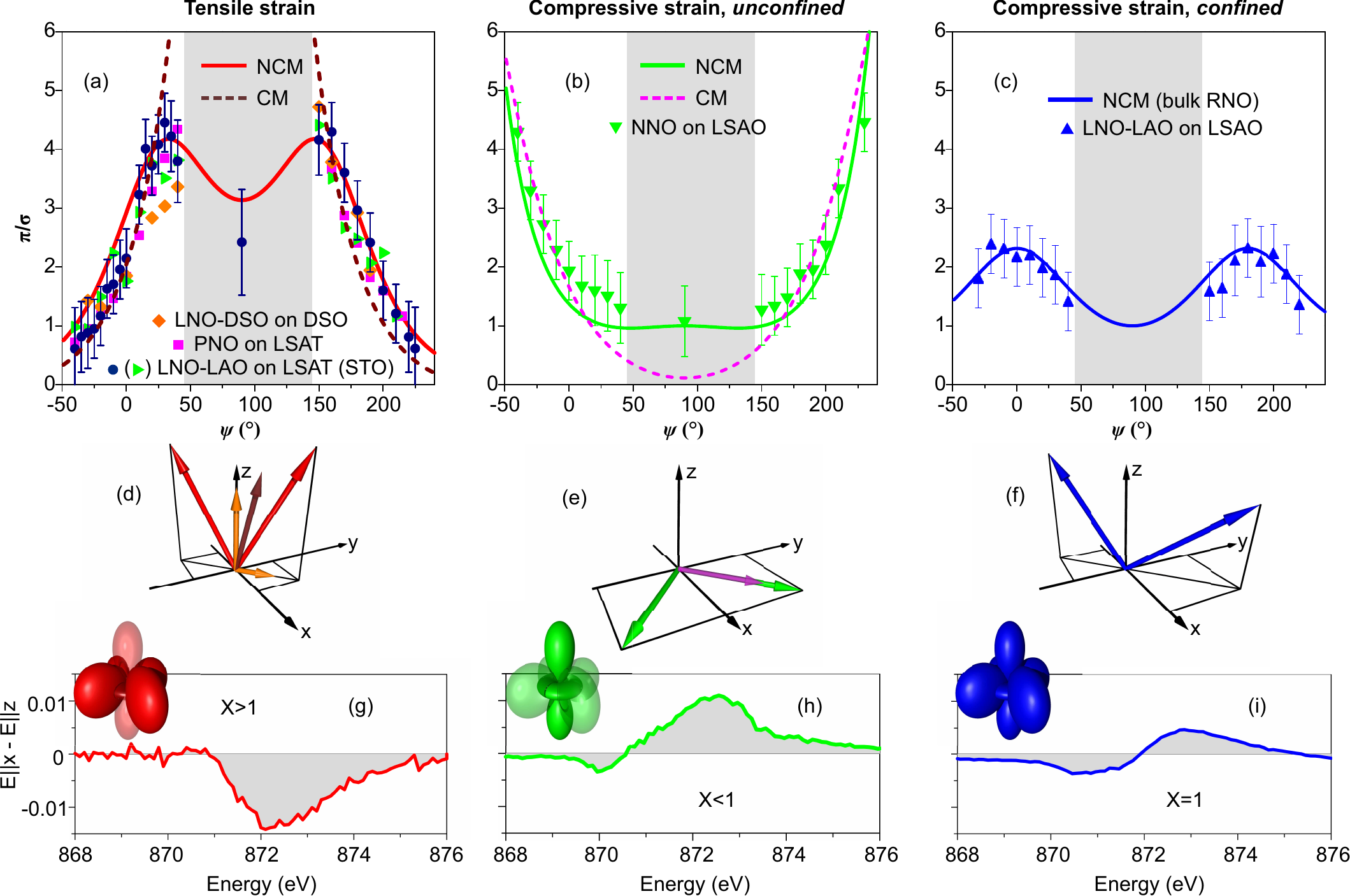}
\caption{(a)-(c) Dependence of the polarization dependent scattering intensity $\boldsymbol\pi /\boldsymbol\sigma$ on the azimuthal angle $\psi$ for LNO-based SLs and a PNO film on LSAT (tensile strain), (b) a NNO film on LSAO (compressive strain, unconfined), and (c), a LNO-LAO SL on LSAO (compressive strain, confined). The solid and dashed lines represent the results of fits to non-collinear (NCM)  and collinear (CM) magnetic structure models described in the text, respectively. In the shaded areas around $\psi=90^{\circ}$, measurements were not possible in the tilted scattering geometry of Fig.~\ref{fig:figure1}, and an alternative geometry was used (see text). (d)-(f) Sketches of the spin directions in the two-sublattice magnetic structure resulting from the fits. The color-coding refers to panel (a)-(c)  \cite{note}. (g)-(i) Linear dichroism obtained from XAS measurements around the Ni $L_2$ edge with incoming light polarization in and out of the substrate plane. The ratio $X$ of orbital occupations derived from the sum rule (see text) is given in the legend and schematically depicted by color-coded solid and semitransparent plots of the corresponding electron density distribution.}
\label{fig:figure3}
\end{figure*}

In order to determine the SDW polarization, we systematically varied the incoming and outgoing photon polarization vectors, $\varepsilon$ and $\varepsilon'$, relative to the electronic magnetic moments by scanning the azimuthal angle $\psi$ around $\textbf{q}_{SDW}$ (Fig.~\ref{fig:figure1}). The intensity of a magnetic Bragg reflection is given by \cite{Haverkort2010}
\begin{equation}\label{scattering}
I_{kl}= \mid\sum_{j} e^{i\textbf{q} \cdot \textbf{r}_j} \varepsilon'_{l} \cdot F_j(E) \cdot \varepsilon_{k}\mid^2,
\end{equation}
with $k,l\in\{\sigma,\pi\}$. Figure~\ref{fig:figure2}(b) shows that the energy profiles of the Bragg reflections measured at different azimuthal angles $\psi$ are identical. This confirms that the Bragg intensity at $\textbf{q}_{SDW}$ arises from a single set of scattering tensors $F_j(E)$ at lattice site $j$, which (together with the corresponding phase factors) encode the magnetic ordering pattern. We measured the scattered intensity for $\sigma$- and $\pi$-polarized incident photons (Fig.~\ref{fig:figure1}), so that the two channels $\boldsymbol\pi\equiv I_{\pi \sigma}+I_{\pi \pi}$ and $\boldsymbol\sigma\equiv I_{\sigma\pi}$ were distinguished. We present the measured data as the ratio $\boldsymbol\pi /\boldsymbol\sigma$, which is not influenced by the shape of the sample and the orientation of its surfaces relative to the incoming and outgoing x-ray beams, and can thus be directly compared to model calculations.

Figure \ref{fig:figure3} displays the polarization-dependent data for different LNO-based SLs with $N=2$ and films exhibiting SDW order, grown on substrates that impose either tensile (Fig.~\ref{fig:figure3}(a)) or compressive (Figs.~\ref{fig:figure3}(b,c)) strain. The blue line in Fig.~\ref{fig:figure3}(c) represents the result of a calculation for the magnetic structure of bulk $R$NiO$_3$ with $R \neq$ La \cite{Garcia1992,Scagnoli2006}, which comprises two sets of spirals polarized perpendicular to the propagation vector $\textbf{q}_{SDW}$ (blue arrows in Figs.~\ref{fig:figure3}(f)). Whereas the magnetic structure of a LNO-LAO SL under compressive strain is compatible with the one in the bulk (Fig.~\ref{fig:figure3}(c)), the data sets on all samples grown under tensile-strain conditions (Fig.~\ref{fig:figure3}(a)), and the one on a NNO film grown under compressive strain (Fig.~\ref{fig:figure3}(b)), indicate distinctly different magnetic structures.

In analyzing the azimuthal scans of Fig.~\ref{fig:figure3}, we considered both collinear antiferromagnetic structures with spatial variation of the moment amplitude and spiral structures analogous to the one in bulk $R$NiO$_3$ with identical amplitude on every lattice site (see Fig.~\ref{fig:figure3}(a,b)). Model calculations \cite{LeePRL2011,LeePRB2011} show that the relative stability of these two structures depends on factors that are difficult to compute from first principles, such as the on-site correlation strength. The comprehensive data sets displayed in Fig.~\ref{fig:figure3} allow an experimental test of these predictions. Whereas collinear structures turned out to be incompatible with the data (see dashed lines in Fig.~\ref{fig:figure3}(a,b)), calculations based on spiral SDWs yield excellent descriptions of all three distinct data sets (solid lines in Figs.~\ref{fig:figure3}(a,b))). These structures can be derived from the structure of bulk $R$NiO$_3$ by adjusting the direction of the moments in the two sublattices. In the best fits for all samples under tensile strain, the moments are symmetrically tilted from the $[001]$-axis by $28\pm2^{\circ}$ (red arrows in Fig.~\ref{fig:figure3}(d)), and remain coplanar with those of the bulk structure \cite{note}. The magnetic structure of the NNO sample under compressive strain, on the other hand, comprises moments along $[110]$ and $[1\overline{1}0]$ (green arrows in Fig.~\ref{fig:figure3}(e)). Note the striking difference between the SDW polarization of this sample and the one of the LNO-LAO SL shown in Fig.~\ref{fig:figure3}(f), which was grown on the same substrate (LSAO) and exhibits the same in-plane lattice parameter.

In order to uncover the origin of the surprising variability of the SDW polarization, we used x-ray absorption spectroscopy with linearly polarized x-rays near the Ni $L_2$ edge to determine the relative occupation of the Ni $d$-orbitals, which controls the magneto-crystalline anisotropy via the spin-orbit coupling \cite{Bruno1989,Csiszar2005}. Figs.~\ref{fig:figure3}(g-i) display the difference of the absorption spectra for photons polarized parallel and perpendicular to the substrate surface for three representative samples. Taking advantage of the sum rule for linear dichroism, we have converted the energy-integrals into the ratio of $e_g$ hole occupation numbers $X\equiv\underline{n}_{3z^2-r^2}/\underline{n}_{x^2-y^2}$ \cite{vanderLaan1994}. The SLs and films grown under tensile strain show $1.03 \leq X \leq 1.14$, corresponding to an enhanced electron occupation of the $d_{x^2-y^2}$ orbital (Fig.~\ref{fig:figure3}(g)). Since orbital moments in this situation will point along $z$, the experimentally observed canting of the spin moments towards this direction (Fig.~\ref{fig:figure3}(d)) is a natural consequence of the intra-atomic spin-orbit coupling. Conversely, the preferential $d_{3z^2-r^2}$ occupation ($X = 0.97$) found in the compressively strained NNO film (Fig.~\ref{fig:figure3}(h)) accounts for the observed spin polarization in the $x,y$-plane (Fig.~\ref{fig:figure3}(e)). In the compressively strained LNO-LAO SL, our data show that the equal population of $d_{x^2-y^2}$ and $d_{3z^2-r^2}$ found in bulk nickelates is restored (Fig.~\ref{fig:figure3}(i)), consistent with prior reports on related SLs \cite{Freeland2011} and with the bulk-like magnetic structure of this sample (Fig.~\ref{fig:figure3}(f)). This requires a mechanism that counteracts the effect of compressive strain observed in the NNO film on the same substrate. Such a mechanism has been identified in model calculations for LNO-based SLs, \cite{Chaloupka2008,Hansmann2009,Han2010} which indicate that the confinement of the LNO conduction electrons to the $x,y$-plane by the insulating blocking layers of the SL stabilizes the $d_{x^2-y^2}$ orbital relative to $d_{3z^2-r^2}$. In contrast to the long-range effect of epitaxial strain, recent results from resonant x-ray reflectometry \cite{Benckiser2011,Wu2013} indicate that the orbital polarization is spatially modulated due to the dimensional confinement which predominantly acts on the interface layers, thus affording an independent means of controlling the SDW polarization.

In summary, our RXD data have allowed us to develop a detailed picture of a complex in-plane magnetic superstructure in an oxide superlattice. Along with recent neutron diffraction work on manganate superlattices \cite{Santos}, this represents one of few cases where such a comprehensive description has proven possible. Note that neutron diffraction is not applicable to systems with few magnetically ordered atomic layers, such as our SLs and most other oxide heterostructures. Our demonstration that RXD experiments are feasible thus indicates new perspectives for this field of research. We have further demonstrated understanding and control of the noncollinear spin polarization by virtue of two independent mechanisms, epitaxial strain and quantum confinement. Since the nickelates can be epitaxially integrated with other metallic oxides exhibiting ferromagnetic or superconducting order, the control options we have identified provide interesting new perspectives for spintronic devices. In view of the substantial conductance in the SDW state, these may eventually include devices with electronically active antiferromagnetic elements \cite{MacDonald}.

We acknowledge fruitful discussions with G. Khalliulin, Y. Matiks, G. Sawatzky, and R. Scherwitzl, and the financial support from the Deutsche Forschungsgemeinschaft within the framework of the TRR80, project C1.


\begin{thebibliography}{99}
\bibitem{note} For tensile strain, our simulation yields an identical solution for a structure with spin sublattices lying along [001] and [110] and magnetic moment amplitudes that differ by a factor $\sim 2$ (orange arrows in Fig.~\ref{fig:figure3}(d)). Additional work is required to elucidate whether the charge disproportionation between neighboring sites implied by the different moment amplitudes is present in our SLs.
\end{thebibliography}
\end{document}